\documentclass[twocolumn,notitlepage,footinbib,superscriptaddress]{revtex4-1}

\usepackage{graphicx}
\usepackage{amsmath,amssymb,mathtools,bbold}
\usepackage{color}
\usepackage[colorlinks=true,linkcolor=blue,citecolor=blue]{hyperref}
\usepackage{soul}

\newcommand{\rev}[1]{\textcolor{black}{#1}}
\newcommand{\revtwo}[1]{\textcolor{black}{#1}}

\begin{document}

\title{Can nonlinear parametric oscillators solve random Ising models?}

\author{Marcello Calvanese Strinati}
\affiliation{Department of Physics, Bar-Ilan University, 52900 Ramat-Gan, Israel}
\affiliation{\revtwo{Dipartimento di Fisica, Universit\`a di Roma ``La Sapienza'', Piazzale Aldo Moro 5, I-00185 Rome, Italy}}
\author{Leon Bello}
\affiliation{Department of Physics and QUEST Center of Quantum Science and Technology, Bar-Ilan University, 52900 Ramat-Gan, Israel}
\author{Emanuele G. Dalla Torre}
\affiliation{Department of Physics, Bar-Ilan University, 52900 Ramat-Gan, Israel}
\author{Avi Pe'er}
\affiliation{Department of Physics and QUEST Center of Quantum Science and Technology, Bar-Ilan University, 52900 Ramat-Gan, Israel}

\date{\today}

\begin{abstract}
We study large networks of parametric oscillators as heuristic solvers of random Ising models. In these networks, known as coherent Ising machines, the model to be solved is encoded in the coupling between the oscillators, and a solution is offered by the steady state of the network. This approach relies on the assumption that mode competition steers the network to the ground-state solution of the Ising model. By considering a broad family of frustrated Ising models, we show that the most-efficient mode does not correspond generically to the ground state of the Ising model. We infer that networks of parametric oscillators close to threshold are intrinsically not Ising solvers. Nevertheless, the network can find the correct solution if the oscillators are driven sufficiently above threshold, in a regime where nonlinearities play a predominant role. We find that for all probed instances of the model, the network converges to the ground state of the Ising model with a finite probability.
\end{abstract}


\maketitle

\textit{\underline{Introduction}.} Solving large-scale optimization problems has been a quest of uttermost importance during the last decades. In addition to physics, optimization problems are central in several fields of modern science such as finance~\cite{gilli2011numerical}, life science~\cite{zhanglifecience}, biophysics and bioinformatics~\cite{naturecombinatorial}, and artificial intelligence~\cite{ohzechiarticifialintellingence}. Many of these problems belong to the non-deterministic polynomial (NP-hard) complexity class~\cite{karpcomputercomputation}: The computational time required to reach the optimal solution scales exponentially with the size of the problem, making the search for the exact solution often unfeasible using conventional computers, even for problems of realistic sizes.

A viable route to tackle some NP-hard problems is offered by the possibility of mapping them onto classical Ising models~\cite{10.3389/fphy.2014.00005}. Solving the original NP-hard problem translates into finding the ground-state (GS) configuration of the corresponding Ising Hamiltonian, which is also \rev{an} NP-hard task~\cite{Barahona_1982}. A number of heuristic algorithms have been developed \rev{for} efficiently finding solutions for the Ising model, \rev{at least approximately}. Notable examples include the Metropolis algorithm~\cite{metropolis1953}, simulated annealing~\cite{kirkpatrick1983}, and quantum annealing~\cite{PhysRevE.58.5355,Santoro2006,boixo2014}. In the last years, networks of coupled parametric oscillators (POs) have emerged as a novel promising heuristic Ising solver~\cite{PhysRevA.88.063853,marandi2014cim,takata2016cubic,nphoton.2016.68,hamerlyfristratedchain2016,PhysRevA.96.043850,Wang2017,Inagaki603,1805.05217,PhysRevLett.122.213902,10.1007/978-3-030-19311-9_19,Pierangeli:20}. This platform, called parametric-oscillator coherent Ising machine (PO-CIM), simulates the dynamics of coupled artificial Ising spins to efficiently find the GS of the corresponding Ising model, specified by the coupling matrix of the network.

In a PO-CIM, below the oscillation threshold, the POs are in a squeezed vacuum state~\cite{PhysRevA.29.408,PhysRevA.30.1386,JOSAB.4.001465, Lvovsky_2015}. Above threshold, the POs undergo a series of pitchfork \revtwo{bifurcations}~\cite{strogatz2007nonlinear}: For a given coupling matrix defining the \emph{graph} of the system, mode competition selects the most efficient mode of the network, i.e., the mode that minimizes the overall loss/gain ratio. In this mode, each oscillator has a binary phase~\cite{landau1982mechanics,strogatz2007nonlinear}. The \rev{working assumption} of PO-CIMs is that the most efficient mode, in terms of phases, is given by the GS configuration of the corresponding Ising model~\cite{PhysRevA.88.063853,marandi2014cim,takata2016cubic,nphoton.2016.68,Inagaki603,1805.05217}.

Here, we show that this naive explanation is insufficient to capture the true working principle of PO-CIMs. By considering a paradigmatic family of random graphs, we demonstrate that the most efficient mode of the network, \rev{which determines the steady state} at threshold, generically does not correspond to the Ising solution. This implies that \rev{POs} at threshold do not solve the desired Ising problem. \rev{Next, we consider the dynamics far above threshold by means of a low-level numerical simulation that closely mimics the temporal dynamics within a network of coupled POs~\cite{Calvanese_Strinati_2020}. We show that, when each oscillator is pumped individually, the system can flow towards the correct Ising solution in a probabilistic manner.}

\textit{\underline{Linear system}.} We open by discussing \rev{PO networks} in the linear regime, i.e., when the pump power is  close to the oscillation threshold. \rev{In this regime, nonlinearities do not affect the steady-state phase configuration: A PO network  converges to the phase configuration dictated by the eigenvector of the coupling matrix $\mathbf{C}$ that corresponds to the eigenvalue with maximal real part~\cite{hamerlyfristratedchain2016} (denoted respectively as the most-efficient eigenvector and eigenvalue hereon). Here, we show that this eigenvector configuration often is not the Ising solution, and thus conclude that a PO network near threshold does not work as an Ising solver.}

To achieve this goal, we denote the steady-state configuration of the network by $\mathbf{\overline A}=(\overline{A}^{(1)},\ldots\overline{A}^{(N)})$, where \rev{$N$ is the number of POs, and} the real number $\overline{A}^{(j)}$ is the amplitude of the $j$-th oscillator (in the rotating frame of the carrier). The overline denotes the steady-state value. We next identify each oscillator with an Ising variable $\sigma={\rm sgn}\mathbf{\overline{A}}$ and check \rev{whether} this configuration minimizes the Ising energy
\begin{equation}
E(\{\sigma_j\})=-\frac{1}{2}\sum_{j,k}C_{jk}\sigma_j\sigma_k \,\, .
\label{eq:isingenergy}
\end{equation}
\rev{In general, the coupling between any two oscillators $j$ and $k$ is written in terms of an energy-preserving (anti-symmetric) part $r_{jk}=-r_{kj}$, and a dissipative (symmetric) part $\alpha_{jk}=\alpha_{kj}$, with respect to the exchange $j\leftrightarrow k$~\cite{PhysRevLett.123.083901,PhysRevA.100.023835,Calvanese_Strinati_2020}: $C_{jk}=r_{jk}+\alpha_{jk}$. The energy-preserving part induces a coherent exchange of energy between the oscillators, while the dissipative part gives rise to an Ising coupling~\cite{hamerlyfristratedchain2016}.} Before proceeding, it is important to discuss what are the possible \rev{classes of} behaviours \rev{for} the system at threshold. Recently, it was found that small PO networks at threshold can \rev{be} significantly \rev{different} from coupled Ising spins. Indeed, the system can show dynamics beyond the Ising picture, such as persistent coherent beats~\cite{PhysRevLett.123.083901,PhysRevA.100.023835} or POs that remain in the zero-amplitude state \rev{also} above threshold~\cite{Calvanese_Strinati_2020}, where the Ising variable is not defined.

\begin{figure}[t]
\centering
\includegraphics[width=8.8cm]{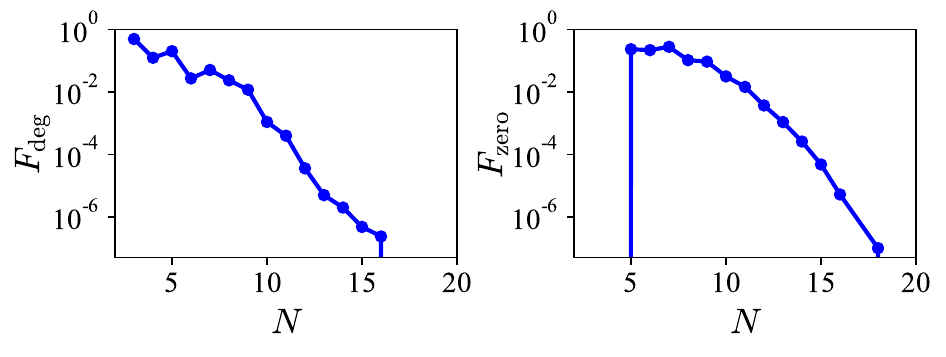}
\caption{Finite-size scaling of $F_{\rm deg}$ (left panel) and $F_{\rm zero}$ (right panel). The statistics was performed by considering the totality of the graphs for $N\leq7$, and by randomly sampling $10^7$ SK graphs for $N>7$. Within our numerical precision, we find that, for $N>16$, $100\%$ of the SK graphs is non-degenerate.}
\label{fig:finitesizescalingnondegenerategraphs}
\end{figure}

Persistent coherent beats at threshold are generically observed when the most-efficient eigenvalue of the coupling matrix is degenerate (\emph{degenerate graphs}). In this case, the steady-state configuration of the system is a linear combination of the most-efficient eigenvectors $\{\mathbf{v}_{{\rm max},m}\}$ of $\mathbf{C}$: $\mathbf{\overline A}=\sum_mc_m\mathbf{v}_{{\rm max},m}$, for some coefficients $\{c_m\}$ determined by the initial conditions. \rev{When $r_{jk}\neq0$}, the network periodically explores all the possible phase configurations admitted by the linear combination, and beats are found~\cite{PhysRevLett.123.083901,PhysRevA.100.023835,Calvanese_Strinati_2020}. \rev{Instead, for} purely dissipative coupling \rev{($r_{jk}=0$)}, the network converges to a specific \rev{phase} configuration $\mathbf{\sigma}={\rm sgn} \mathbf{\overline A}$ admitted by the linear combination. If $\mathbf{\sigma}$ minimizes the Ising energy in Eq.~\eqref{eq:isingenergy}, the network at threshold always succeeds to find the Ising solution. Otherwise, it always fails. 

If the most-efficient eigenvalue is non-degenerate, two possible cases are found: Either the entries of the most-efficient eigenvector $\mathbf{v}_{\rm max}$ are all nonzero (\emph{non-degenerate graphs}), or some of them are zero (\emph{zero graphs}). For zero graphs, the POs that correspond to the zero entries of $\mathbf{v}_{\rm max}$ remain in the zero-amplitude state \rev{also} above threshold. In this case, the network intrinsically fails to behave as a PO-CIM, since POs with zero amplitudes do not represent a valid Ising solution. Instead, for non-degenerate graphs, the network at threshold converges to two possible Ising configurations only ($\sigma=\pm{\rm sgn}\mathbf{v}_{\rm max}$). If $\mathbf{\sigma}$ minimizes Eq.~\eqref{eq:isingenergy}, the system converges to the correct Ising solution, otherwise it does not. Notable examples of graphs where the PO network finds correct Ising solutions were reported in~\cite{marandi2014cim,takata2016cubic,kalinin2020}.

\begin{figure}[t]
\centering
\includegraphics[width=7.9cm]{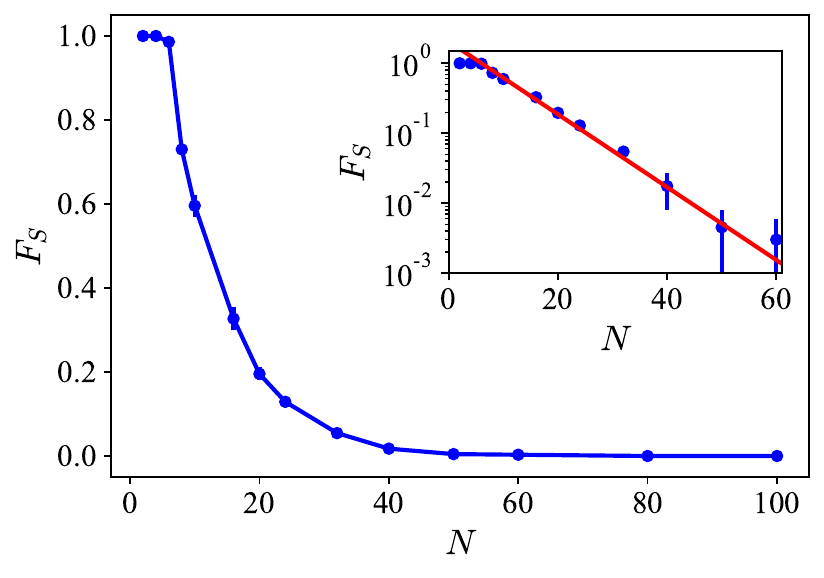}
\caption{Finite-size scaling of $F_S$ obtained by randomly sampling $300$ non-degenerate SK graphs. We find $F_S=100\%$ for $N<6$, and it exponentially decreases for $N\geq6$, as highlighted by the exponential fit (red solid line) in the inset.}
\label{fig:finitesizescalingfailedgraphs}
\end{figure}

\rev{The discussion above reviewed the variety of behaviours of coupled POs. To specifically discuss PO networks as Ising solvers}, we now \rev{focus on} Sherrington-Kirkpatrick (SK) graphs, \rev{with} symmetric coupling matrices $C_{jk}=C_{kj}$ (i.e., purely dissipative coupling) \rev{that} represent a valid Ising graph. Specifically, we take $C_{jk}=\pm1$ with $\pm1$ randomly chosen with equal probability.

We first analyze the fraction of zero and degenerate graphs, respectively $F_{\rm zero}$ and $F_{\rm deg}$, as a function of $N$. The result is shown in Fig.~\ref{fig:finitesizescalingnondegenerategraphs}. With our numerical resources, due to the exponentially increasing number of SK graphs as a function of $N$ (there are $2^{N(N-1)/2}$ \rev{possible} SK graphs for a given $N$), we can systematically analyze all graphs only up to $N=7$. Thus, for $N>7$, we randomly sample $10^7$ graphs and assume this to be a fair sample of the total graph population. As evident from the figure, $F_{\rm deg}$ and $F_{\rm zero}$ quickly decay to zero, indicating that the probability of finding zero or degenerate graphs becomes vanishingly small as $N$ is increased. The result in Fig.~\ref{fig:finitesizescalingnondegenerategraphs} allows us to focus from now on only on non-degenerate SK graphs.

Next, we analyze if the steady state of the network at threshold is the correct Ising solution. To do so, we proceed as follows: For different values of $N$, we randomly select a number $W$ of non-degenerate SK graphs. For each graph, we compute the most-efficient eigenvector $\mathbf{v}_{\rm max}$ and the corresponding Ising energy $E(\{\sigma_j\})$ as in Eq.~\eqref{eq:isingenergy} with $\mathbf{\sigma}={\rm sgn}\mathbf{v}_{\rm max}$. Then, we compare the energy computed from $\mathbf{v}_{\rm max}$ with the GS energy $E_{\rm GS}$ of the corresponding Ising model. If $E(\{\sigma_j\})=E_{\rm GS}$, we conclude that the network at threshold \rev{works as a PO-CIM} for that graph.

We randomly select $W=300$ non-degenerate SK graphs and compute the \emph{success fraction} $F_S(N)$ defined as the fraction of graphs for which the network at threshold finds a correct Ising solution. To estimate the numerical precision of our algorithm, we repeat the same procedure five times and plot in Fig.~\ref{fig:finitesizescalingfailedgraphs} the average and standard deviation of the results obtained by this technique. We find that the network at threshold works as a PO-CIM for small network sizes $N<6$, but for larger values of $N$, the success fraction exponentially decays, as highlighted by the exponential fit in the inset. The data for $N=80,100$ are absent from the logarithmic plot since we find $F_S=0$ within our numerical precision. The unavoidable conclusion is that the success probability of a PO network near threshold to find the Ising \rev{solution} of a \rev{large} random graph is exceedingly small.

\vspace{0.2cm}
\textit{\underline{Nonlinear system}.} We now extend the previous analysis to the nonlinear regime by computing the success fraction when the pump power is above the oscillation threshold. Our goal is to show that, even if the network fails to behave as PO-CIM at threshold, nonlinear effects can induce the network to find \rev{the} correct Ising solution sufficiently above threshold. We show how this fact critically depends on the \rev{exact} form of the nonlinearity.

\begin{figure}[t]
\centering
\includegraphics[width=7.5cm]{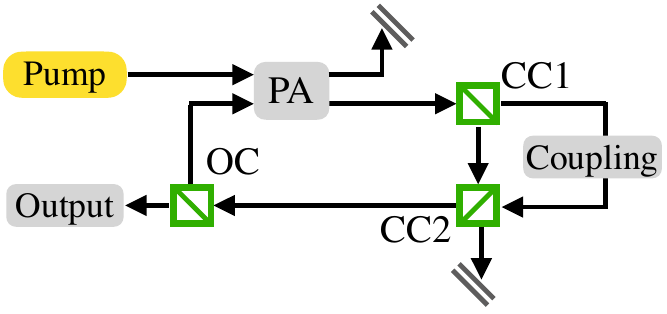}
\caption{Block scheme of the low-level numerical simulation~\cite{Calvanese_Strinati_2020}. The cavity consists of a parametric amplifier (PA) where the signal and the pump field are injected, two coupling couplers (CC) that connect the cavity to the coupling device, and an output coupler (OC) to extract part of the signal from the cavity and send it to the measurement apparatus.}
\label{fig:simulatedexperimentscheme}
\end{figure}

To analyze the system above threshold, we simulate the dynamics of the network by means of the low-level numerical simulation in~\cite{Calvanese_Strinati_2020} (see Fig.~\ref{fig:simulatedexperimentscheme}). At a given round trip $n\in~\mathbb{N}$, the signal consists of a $N$-dimensional complex vector $\mathbf{A}_n$. \rev{To take into account the effects of the noise, we assign initial conditions $\mathbf{A}_0$ randomly selected from a uniform distribution with positive and negative values}. The signal is injected together with a pump field of power $h$ into a parametric amplifier (PA), which amplifies the real parts of the fields ${\rm Re}(\mathbf{A}_{n})$ and suppresses the imaginary parts ${\rm Im}(\mathbf{A}_{n})$ following the appropriate nonlinear wave equation~\cite{boyd2008nonlinear,footnotesupplementalmaterial}.

The amplified signal is sent into the coupling device, which is connected to the cavity by two couplers (CC1 and CC2 in Fig.~\ref{fig:simulatedexperimentscheme}) that split the input field according to the transmission and reflection coefficients $T_{c,m}$ and $R_{c,m}=\sqrt{1-T_{c,m}^2}$, with $m=1,2$ for CC1 and CC2, respectively. We take the coupling matrix as $C_{jk}=\pm\alpha$, where $\alpha>0$ is the strength of the dissipative coupling and the signs are randomly chosen with equal probability. While the value of $\alpha$ is irrelevant for the linear analysis, it affects the behaviour of the system in the nonlinear regime, as shown later on. The coupled field is reinjected into the cavity, and eventually the signal is sent to an output coupler (OC), with transmission and reflection coefficients $T_{\rm out}$ and $R_{\rm out}$, and then measured. The \rev{evolution of the fields inside the cavity is captured by the nonlinear map}
\begin{equation}
A^{(j)}_{n+1}=R_{\rm out}\sum_kQ_{jk}{\rm PA}[\{A^{(k)}_{n}\}] \,\, .
\label{eq:equationsforthefiledamplitude11}
\end{equation}
Here, we define $\mathbf{Q}=a\mathbb{1}+b\mathbf{C}$, where $\mathbb{1}$ is the identity matrix, $a=R_{c,1}R_{c,2}$, $b=T_{c,1}T_{c,2}$, and ${\rm PA}[\{A^{(j)}_{n}\}]$ denotes the fields after the parametric amplification~\cite{footnotesupplementalmaterial}.

The dynamics of the network above threshold crucially depends on the nonlinearity, encoded in ${\rm PA}[\{A^{(j)}_{n}\}]$ in Eq.~\eqref{eq:equationsforthefiledamplitude11}. Here, we focus on nonlinearities due to pump depletion, which is the most relevant process in many experimental contexts. We consider two common cases: (i) The case when all POs are pumped by the same pump:
\begin{equation}
{\rm PA}[\{A^{(j)}_{n}\}]\!=\!A^{(j)}_{n}\!+\!\left[gh\!-\!2g^2\sum_k{\left(A^{(k)}_{n}\right)}^2\right]\!\left(A^{(j)}_{n}\right)^* \, ,
\label{eq:equationsforthefiledamplitude1}
\end{equation}
where the star denotes complex conjugation, and (ii) the case when each PO is independently driven by its own pump:
\begin{equation}
{\rm PA}[\{A^{(j)}_{n}\}]\!=A_{n}^{(j)}\!+\!\!\left[gh\!-\!2g^2{\left(A^{(j)}_{n}\right)}^2\right]\!\left(A_{n}^{(j)}\right)^* \,\, ,
\label{eq:equationsforthefiledamplitude2}
\end{equation}
where $g$ in Eqs.~\eqref{eq:equationsforthefiledamplitude1} and~\eqref{eq:equationsforthefiledamplitude2} \rev{quantifies} pump depletion within the nonlinear medium.

The case of pump depletion described in Eq.~\eqref{eq:equationsforthefiledamplitude1} can be treated analytically. In the steady state, one has $A^{(j)}_{n+1}=A^{(j)}_{n}$, which from Eq.~\eqref{eq:equationsforthefiledamplitude11} yields $\mathbf{Q}\mathbf{\overline A}=[(1+g h-2g^2{|\mathbf{\overline A}|}^2)R_{\rm out}]^{-1}\mathbf{\overline A}$. This implies that $\mathbf{\overline A}$ is an eigenvector of the coupling matrix, for any pump power $h$. Since the steady state at threshold is given by $\mathbf{v}_{\rm max}$ and $\mathbf{Q}\mathbf{\overline A}$ does not depend on $h$, the steady-state configuration of the network remains proportional to $\mathbf{v}_{\rm max}$ for all $h$. The network with the nonlinearity in Eq.~\eqref{eq:equationsforthefiledamplitude1} is therefore not an Ising solver, \rev{at any pump power}.

The network can \rev{however} operate as a heuristic Ising solver \rev{with} the nonlinearity in Eq.~\eqref{eq:equationsforthefiledamplitude2}. To analyze the behaviour of the network above threshold, we resort to the numerical simulation of the coherent dynamics described above. We randomly select $W=300$ SK graphs and compute the success fraction $F_S$ for different values of $h$ above the threshold $h_{\rm th}$, for a given set of parameters in the simulation. The deviation of $h$ from the threshold is parametrized as $h=h_{\rm th}(1+\Delta h)$. For each graph and $\Delta h$, we repeat the simulation $M=100$ times assigning different random initial conditions $\mathbf{A}_0$ at each repetition. At the end of each repetition $\mu=1,\ldots M$, we \rev{compute} the Ising energy $E_\mu(\{\sigma_j\})$ \rev{of the obtained PO state} from Eq.~\eqref{eq:isingenergy}  ($\mathbf{\sigma}={\rm sgn}\mathbf{\overline A}$), and compare it with the calculated GS energy $E_{\rm GS}$ of the Ising model. Out of $M$ repetitions, the condition $E_\mu(\{\sigma_j\})=E_{\rm GS}$ is found $M_{\rm succ}$ times: If $M_{\rm succ}>0$ the network \rev{may work probabilistically} as a heuristic Ising solver, whereas if $M_{\rm succ}=0$ we conclude that within our numerical precision the network fails to behave as a PO-CIM.

\begin{figure}[t]
\centering
\includegraphics[width=8.8cm]{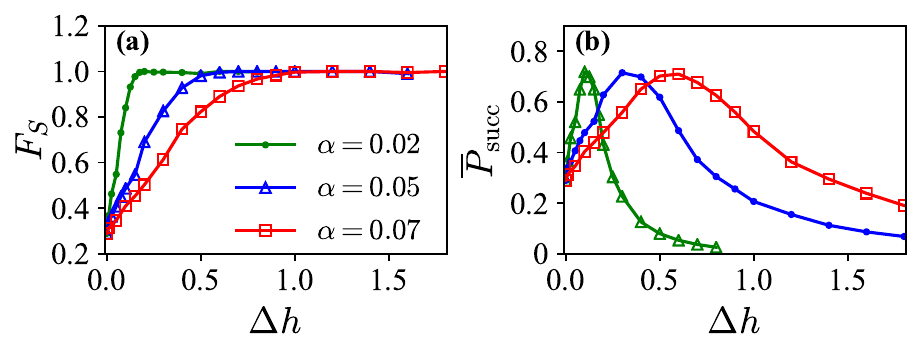}
\caption{\textbf{(a)} Success fraction $F_S$ for $N=16$ and different values of the coupling strength $\alpha$, as a function of the pump strength above threshold $\Delta h$ for the nonlinearity as in Eq.~\eqref{eq:equationsforthefiledamplitude2}. \textbf{(b)} Average success probability $\overline P_{\rm succ}$ as a function of $\Delta h$ computed from the distributions in Fig.~\ref{fig:successfractionnonlineartout0p1differentalpha2}.}
\label{fig:successfractionnonlineartout0p1differentalpha}
\end{figure}

The \rev{results} of the simulation \rev{for a network of size $N=16$ are} shown in Fig.~\ref{fig:successfractionnonlineartout0p1differentalpha}\textbf{(a)}. We \rev{select} $g=0.09$, $T_{c,1}=T_{c,2}=0.2$, $T_{\rm out}=0.1$, and show \rev{the success fraction} $F_S$ as a function of \rev{pump power} $\Delta h$ for different values of \rev{the coupling strength} $\alpha$, as in the legends. We see that the success fraction, starting from the threshold value $F_S\simeq29\,\%$ (Fig.~\ref{fig:finitesizescalingfailedgraphs}), increases monotonically and approaches $F_S=100\%$ upon increasing $\Delta h$~\footnote{We verified that with the pump depletion described in Eq.~\eqref{eq:equationsforthefiledamplitude1}, $F_S$ is constantly equal to the value computed at threshold.}. Also, our data show that smaller values of $\alpha$ enhance the convergence of the network to the correct Ising solution, since the network reaches the nonlinear PO-CIM regime for smaller $\Delta h$ with smaller $\alpha$~\cite{PhysRevA.100.023835}.

\begin{figure}[t!]
\centering
\includegraphics[width=8.8cm]{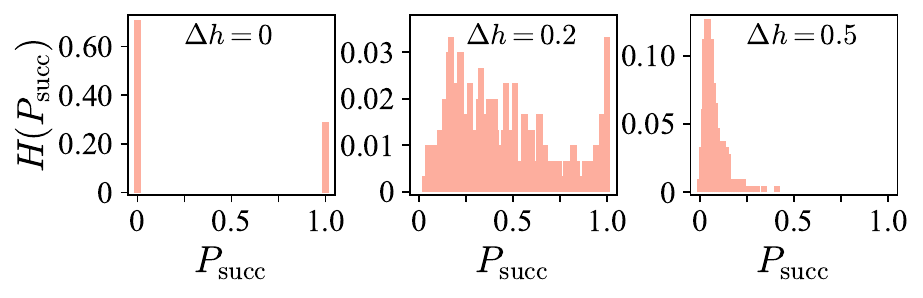}
\caption{Distributions $H(P_{\rm succ})$ of the success probabilities in three prototype cases, computed using $100$ bins, for $\alpha=0.02$ and for $\Delta h$ as in the legends. At threshold ($\Delta h=0$) the distribution is bimodal, and it evolves towards a peaked distribution around $P_{\rm succ}=0$ by increasing $\Delta h$~\cite{footnotesupplementalmaterial}.}
\label{fig:successfractionnonlineartout0p1differentalpha2}
\end{figure}

In addition to $F_S$, a deeper insight on the behaviour of the system is provided by the \emph{success probability}, quantifying how often the \rev{correct} Ising GS is found out of \rev{the} $M$ repetitions \rev{of the simulations}, for a given SK graph and system parameters: $P_{\rm succ}(\mathbf{C})\coloneqq M_{\rm succ}/M$. To compute the success probabilities, we perform a statistical average over the scanned SK graphs as follows: We first compute the distribution of the success probabilities $H(P_{\rm succ})$ \rev{for a given set of parameters}, which quantifies the fraction of \rev{simulated} SK graphs \rev{that exhibit a} success probability $P_{\rm succ}$. \rev{Figure~\ref{fig:successfractionnonlineartout0p1differentalpha2} shows} prototype cases of $H(P_{\rm succ})$ for three different values of $\Delta h$. We checked that the observed behavior is common to all values of $\alpha$~\cite{footnotesupplementalmaterial}. At threshold, $H(P_{\rm succ})$ is bimodal (the SK graphs have either $P_{\rm succ}=0$ or $P_{\rm succ}=1$) because the PO network \emph{deterministically} converges to the configuration dictated by the most-efficient eigenvectors of the coupling matrix, \rev{which either coincides with the Ising GS or does not}. Above threshold, $H(P_{\rm succ})$ starts to be nonzero also for $0<P_{\rm succ}<1$. This implies that, in the nonlinear regime, the PO network \rev{does find} correct Ising solutions, \rev{but} in a \emph{non-deterministic} manner, highlighting the heuristic nature of PO-CIMs. As $\Delta h$ is further increased above threshold, $H(P_{\rm succ})$ becomes peaked around smaller and smaller values of $P_{\rm succ}$.

From the distributions $H(P_{\rm succ})$, we compute the average success probability $\overline{P}_{\rm succ}$~\cite{footnotesupplementalmaterial}, which is shown in Fig.~\ref{fig:successfractionnonlineartout0p1differentalpha}\textbf{(b)}. For $\Delta h=0$, $\overline P_{\rm succ}=F_S$ since $H(P_{\rm succ})$ is bimodal at threshold. As $\Delta h$ is increased above threshold, $\overline P_{\rm succ}$ increases until a maximum value is reached. By further increasing $\Delta h$, $\overline P_{\rm succ}$ decreases towards zero, due to the fact that $H(P_{\rm succ})$ becomes more and more peaked around $\overline P_{\rm succ}=0$ for pump powers high above threshold. This non-monotonic behaviour of $\overline P_{\rm succ}$ reveals that there exists an optimal value of pump power sufficiently above the oscillation threshold that maximizes the efficiency of the PO-CIM.

\vspace{0.2cm}
\textit{\underline{Conclusions}.} To conclude, we reported a detailed analysis of a network of $N$ coupled parametric oscillators (POs), analyzing the regimes of parameters where the network can be used as a coherent Ising machine (PO-CIM). Close to the oscillation threshold, the POs adjust their amplitudes and phases to converge to the minimal-loss configuration of the network, dictated by the eigenvectors of coupling matrix of the system corresponding to the maximal eigenvalue. We considered a paradigmatic family of Ising graphs (the SK model) and showed that the minimal-loss configuration often does not coincide with the ground-state solution of the corresponding Ising model. Instead, if each PO is driven individually, the network finds good Ising solutions with finite success probability in the nonlinear regime of operation, for pump powers sufficiently above the oscillation threshold.

On one hand, our findings show that parametric oscillators networks at threshold are intrinsically not Ising solvers, due to the linearity of the system at threshold. On the other hand, they highlight the role of nonlinear effects in realizing PO-CIMs. Our findings are an important step towards the realization and optimization of networks of coupled POs as heuristic Ising solver. We note that nonlinearities lead to an improvement of the success ratio, analogous to recently proposed feedback mechanisms that force all the POs to have the same amplitude~\cite{PhysRevLett.122.040607,doi:10.1002/qute.202000045,Vadlamani26639}. One intriguing possibility is that the key effect of nonlinearities is the reduction of the amplitude inhomogeneity with respect to the linear solution.

\vspace{0.2cm}
\textit{\underline{Acknowledgements}.} We thank Igal Aharonovich, Geva Arwas, and Nir Davidson for fruitful discussions. We are grateful to Richard Berkovits and Davide Rossini for support. This work was supported by the Israel Science Foundation (ISF) Grants No. 151/19, and 154/19, and US-Israel Binational Science Foundation (BSF) Grants No. 2017743, 2016130, and 2018726.


%

\onecolumngrid
\pagebreak
\begin{center}
\textbf{\large Supplemental Material for ``Can nonlinear parametric oscillators solve random Ising models?''}
\end{center}
\setcounter{equation}{0}
\setcounter{figure}{0}
\setcounter{table}{0}
\setcounter{page}{1}
\makeatletter
\renewcommand{\theequation}{S\arabic{equation}}
\renewcommand{\thefigure}{S\arabic{figure}}

\section{Details on the simulation of parametric amplification}
We here provide some details on the low-level numerical simulation discussed in the main text, specifically showing how the parametric amplification term ${\rm PA}[\{A^{(k)}_{n}\}]$ in Eqs.~(2)-(4) is derived from the nonlinear wave equations. We consider a $\chi^{(2)}$ nonlinear crystal of linear length $L$. Let $x$ be the coordinate along which the field propagates inside the nonlinear crystal. When a strong pump field at frequency $\omega_p$ and with amplitude $A_p$ is injected into the nonlinear medium, the pump field is down converted into a signal and idler fields, respectively at frequency $\omega_s$ and $\omega_i$, where $\omega_p=\omega_s+\omega_i$, and with complex amplitudes $A_s$ and $A_i$. The spatial evolution along the nonlinear crystal of $A_p$, $A_s$, and $A_i$ is described by the following set of nonlinear coupled equations~\cite{boyd2008nonlinear}:
\begin{equation}
\frac{dA_s}{dx}=\frac{2i\omega_s^2d_{\rm eff}}{k_sc^2}\,A_pA_i^*e^{-i\Delta kx} \qquad 
\frac{dA_i}{dx}=\frac{2i\omega_i^2d_{\rm eff}}{k_ic^2}\,A_pA_s^*e^{-i\Delta kx} \qquad
\frac{dA_p}{dx}=\frac{2i\omega_p^2d_{\rm eff}}{k_pc^2}\,A_sA_ie^{i\Delta kx} \,\, ,
\label{eq:equationsforthefiledamplitudes1}
\end{equation}
where $d_{\rm eff}$ is the nonlinear coupling coefficient, $k_p$, $k_s$, and $k_i$ are the momenta of the pump, signal, and idler fields, and $\Delta k=k_s+k_i-k_p$ is the phase mismatch. In the following, we consider degenerate parametric amplification: $A_s=A_i\equiv A$, $\omega_s=\omega_i\equiv\omega$, $k_s=k_i\equiv k$, and $\omega_p=\omega_s+\omega_i\equiv2\omega$. We also assume the phase mismatch to be zero, $\Delta k=0$ ($k_p=2k$). By defining $\kappa=2\omega^2d_{\rm eff}/(kc^2)$, Eq.~\eqref{eq:equationsforthefiledamplitudes1} reduces to
\begin{equation}
\frac{dA}{dx}=i\kappa\,A_pA^* \qquad 
\frac{dA_p}{dx}=2i\kappa\,A^2\,\, .
\label{eq:equationsforthefiledamplitudes2}
\end{equation}
In order to explicitly identify the real and imaginary components of the field $A$ [respectively $A_R={\rm Re}(A)$ and $A_I={\rm Im}(A)$] as the amplified and suppressed quadratures in phase-dependent amplification, respectively, it is convenient to shift the phase of the pump field in Eq.~\eqref{eq:equationsforthefiledamplitudes2} by $-\pi/2$~\cite{PhysRevA.100.023835}:
\begin{equation}
\frac{dA}{dx}=\kappa\,A_pA^* \qquad 
\frac{dA_p}{dx}=-2\kappa\,A^2\,\, .
\label{eq:equationsforthefiledamplitudes3}
\end{equation}
By assuming that the field $A$ does not change considerably during propagation in the nonlinear medium (low gain), one can integrate the equation for $A_p$ in Eq.~\eqref{eq:equationsforthefiledamplitudes3} from $x=0$ to $x=L$ to obtain
\begin{equation}
A_p(L)=A_p(0)-2\kappa L A^2 \,\, ,
\label{eq:equationsforthefiledamplitudes4}
\end{equation}
which describes the depletion of the pump field inside the nonlinear medium (gain-saturation effect). By plugging Eq.~\eqref{eq:equationsforthefiledamplitudes4} into the equation for $A$ in Eq.~\eqref{eq:equationsforthefiledamplitudes3}, one obtains ($A_p(0)\equiv h$)
\begin{equation}
\frac{dA}{dx}=\kappa\left(h-2\kappa LA^2\right)A^* \,\, .
\label{eq:equationsforthefiledamplitudes4}
\end{equation}
By integrating Eq.~\eqref{eq:equationsforthefiledamplitudes4} between $x=0$ and $x=L$ as done for the pump field, one has at the end of the nonlinear crystal the signal field ($g\equiv\kappa L$)
\begin{equation}
A(L)=A+\left(gh-2g^2A^2\right)A^*=\left(1-2g^2{|A|}^2\right)A+ghA^* \,\, ,
\label{eq:equationsforthefiledamplitudes5}
\end{equation}
where it is intended that $A\equiv A(0)$. To make phase-dependent amplification evident, we can separate in Eq.~\eqref{eq:equationsforthefiledamplitudes5} the real and imaginary parts of the field (${|A|}^2=A_R^2+A_I^2$)
\begin{equation}
\left\{\begin{array}{l}
A_R(L)=\left[1+gh-2g^2{|A|}^2\right]A_R\\\\
A_I(L)=\left[1-gh-2g^2{|A|}^2\right]A_I
\end{array}\right. \,\, .
\label{eq:equationsforthefiledamplitudes6}
\end{equation}
Parametric amplification amplifies $A_R$ and suppresses $A_I$. Equation~\eqref{eq:equationsforthefiledamplitudes5} identifies the field amplitude at the end of the nonlinear crystal after single pass inside the nonlinear medium, i.e., at a specific round-trip $n$ of the field inside the parametric cavity. When considering a system of $N$ POs, and each PO amplitude $A^{(j)}_n$ is amplified by its own pump field, then $A_n^{(j)}(L)\equiv {\rm PA}[\{A_n^{(j)}\}]$ denotes the POs amplitudes after the parametric amplification at the $n$-th round trip, which is Eq.~(4) of the main text. When instead the POs are pumped by the same pump field, one can repeat the same calculation but rewriting $dA_p/dx= -2\kappa\sum_{k=1}^{N}{\left(A^{(k)}\right)}^2$, to obtain Eq.~(3) of the main text.

\section{Numerical data of the success probability distributions}
We here show the data of the distributions $H(P_{\rm succ})$ used to compute the average success probability $\overline{P}_{\rm succ}$ in Figs.~4 and~5 of the main text, for all scanned values of $\alpha$. As explained in the main text, we use the numerical parameters $N=16$, $g=0.09$, $T_{c,1}=T_{c,2}=0.2$, and $T_{\rm out}=0.1$. The statistics is performed over $W=300$ randomly chosen SK graphs. By repeating the experiment $M=100$ times for each graph, and by determining the number of repetitions $M_{\rm succ}$ for which the POs network finds corrects Ising solutions, we define the success probability as $P_{\rm succ}=M_{\rm succ}/M$. The distribution $H(P_{\rm succ})$ quantifies the fraction of SK graphs with success probability $P_{\rm succ}$.

\begin{figure}[h]
\flushleft{$\alpha=0.02$}\\\vspace{0.0cm}
\centering
\includegraphics[width=4.4cm]{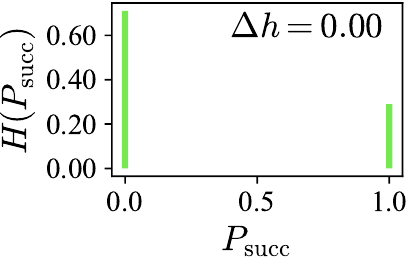}
\includegraphics[width=4.4cm]{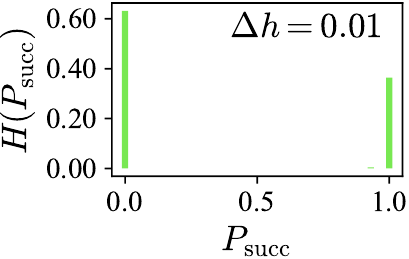}
\includegraphics[width=4.4cm]{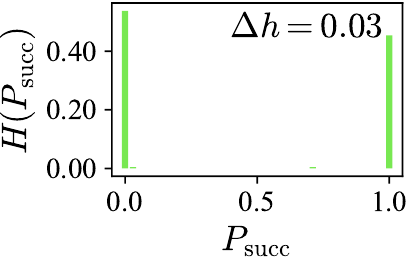}
\includegraphics[width=4.4cm]{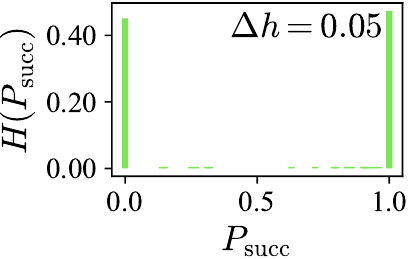}
\includegraphics[width=4.4cm]{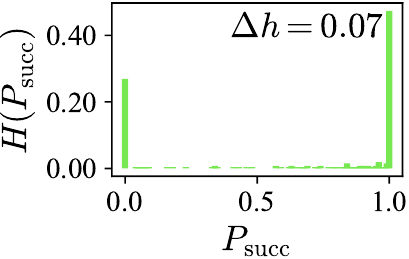}
\includegraphics[width=4.4cm]{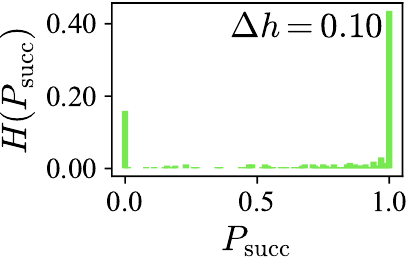}
\includegraphics[width=4.4cm]{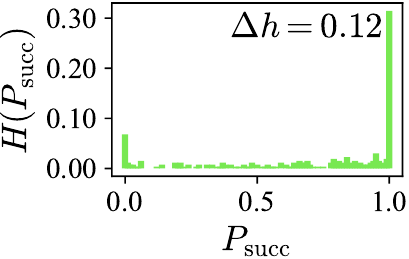}
\includegraphics[width=4.4cm]{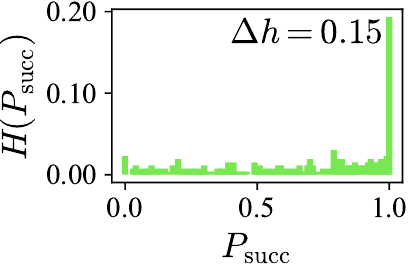}
\includegraphics[width=4.4cm]{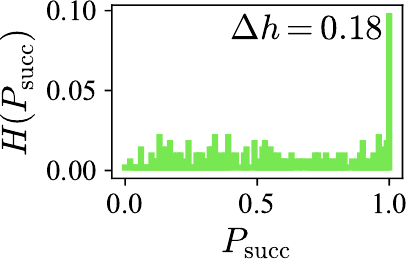}
\includegraphics[width=4.4cm]{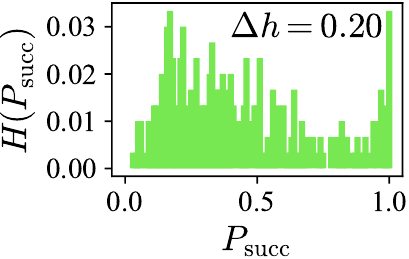}
\includegraphics[width=4.4cm]{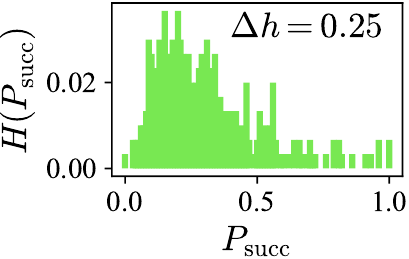}
\includegraphics[width=4.4cm]{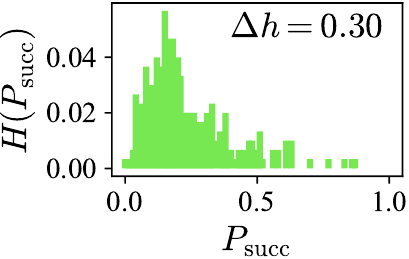}
\includegraphics[width=4.4cm]{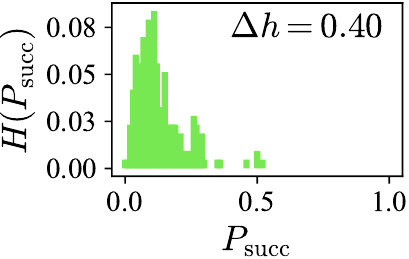}
\includegraphics[width=4.4cm]{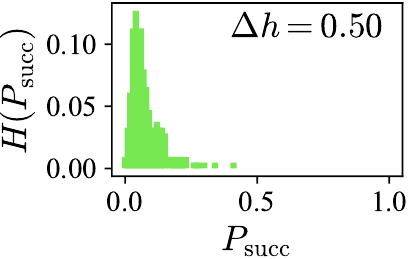}
\includegraphics[width=4.4cm]{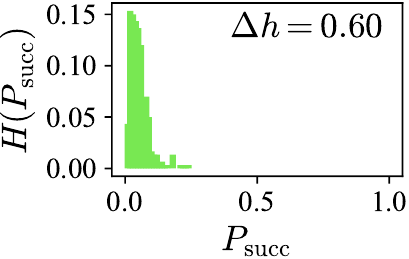}
\includegraphics[width=4.4cm]{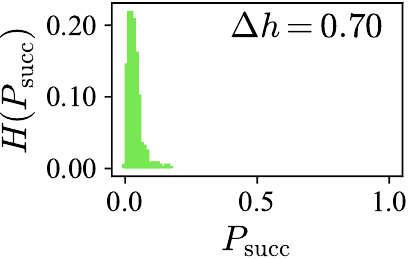}
\includegraphics[width=4.4cm]{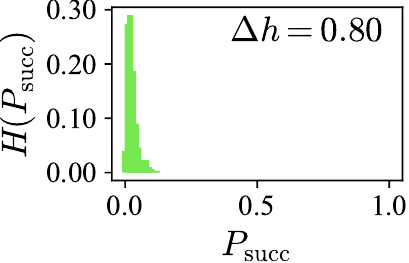}
\caption{Distributions $H(P_{\rm succ})$ for $\alpha=0.02$ and different values of $\Delta h$ as in the legends.}
\label{fig:successprobability1}
\end{figure}

\begin{figure}[h]
\flushleft{$\alpha=0.05$}\\\vspace{0.0cm}
\centering
\includegraphics[width=4.4cm]{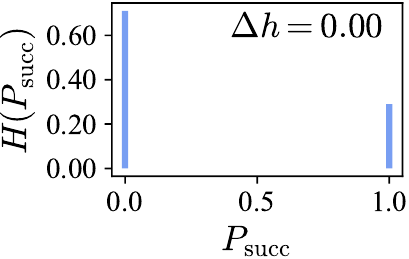}
\includegraphics[width=4.4cm]{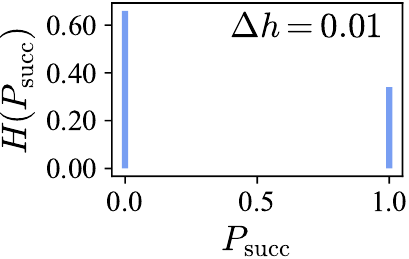}
\includegraphics[width=4.4cm]{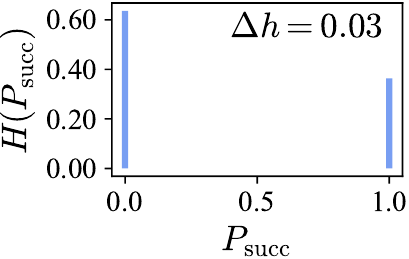}
\includegraphics[width=4.4cm]{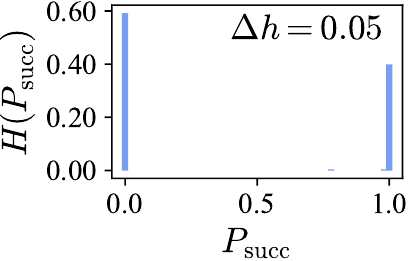}
\includegraphics[width=4.4cm]{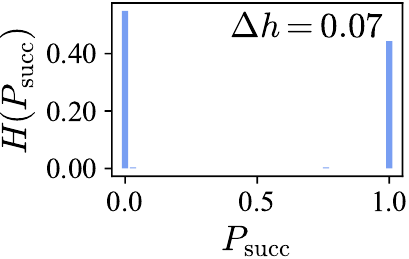}
\includegraphics[width=4.4cm]{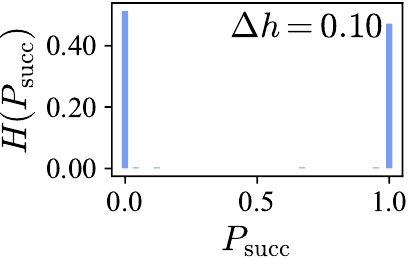}
\includegraphics[width=4.4cm]{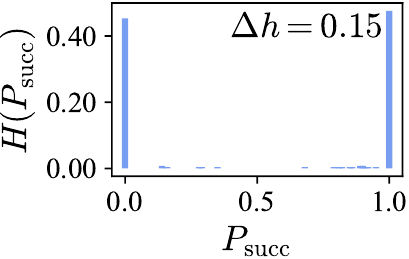}
\includegraphics[width=4.4cm]{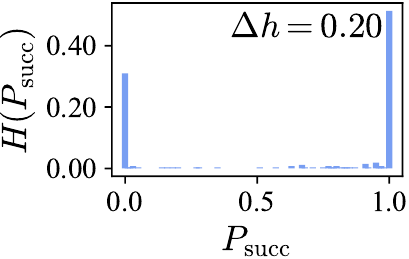}
\includegraphics[width=4.4cm]{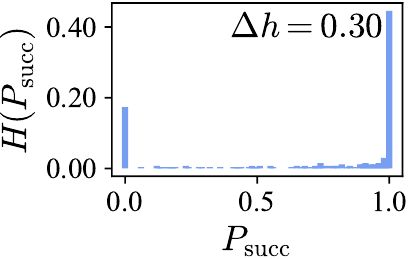}
\includegraphics[width=4.4cm]{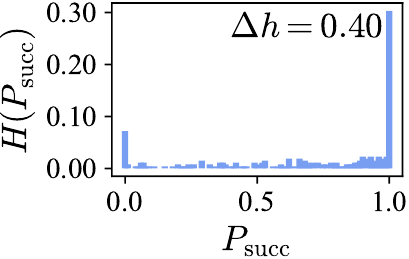}
\includegraphics[width=4.4cm]{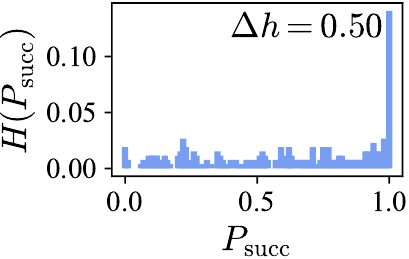}
\includegraphics[width=4.4cm]{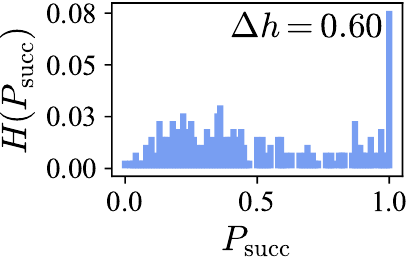}
\includegraphics[width=4.4cm]{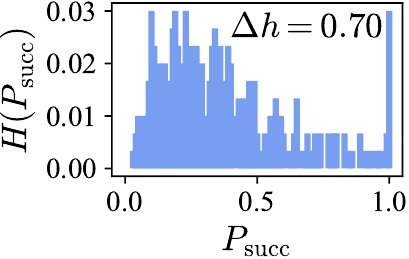}
\includegraphics[width=4.4cm]{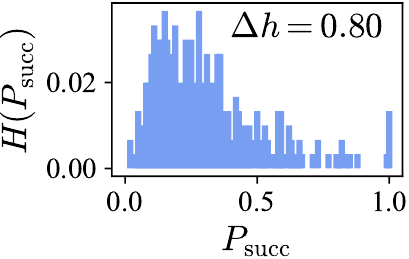}
\includegraphics[width=4.4cm]{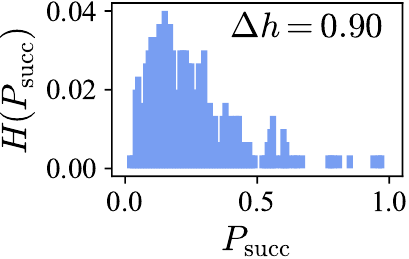}
\includegraphics[width=4.4cm]{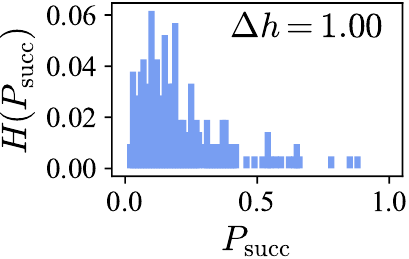}
\includegraphics[width=4.4cm]{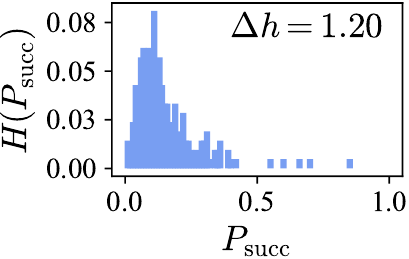}
\includegraphics[width=4.4cm]{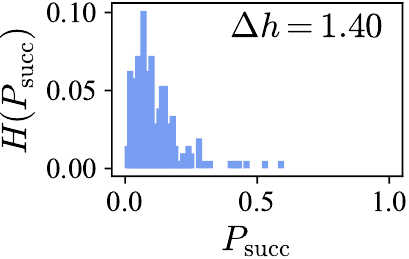}
\includegraphics[width=4.4cm]{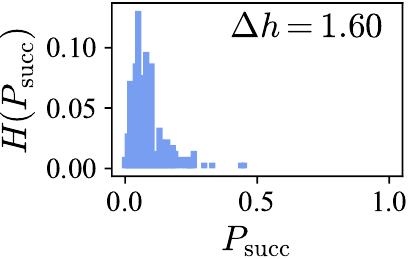}
\includegraphics[width=4.4cm]{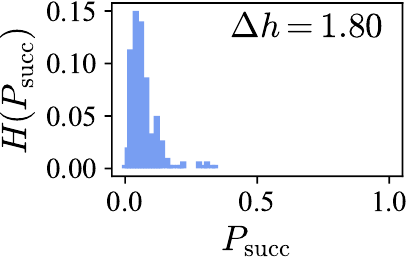}
\caption{Distributions $H(P_{\rm succ})$ for $\alpha=0.05$ and different values of $\Delta h$ as in the legends.}
\label{fig:successprobability2}
\end{figure}

\begin{figure}[h]
\flushleft{$\alpha=0.07$}\\\vspace{0.0cm}
\centering
\includegraphics[width=4.4cm]{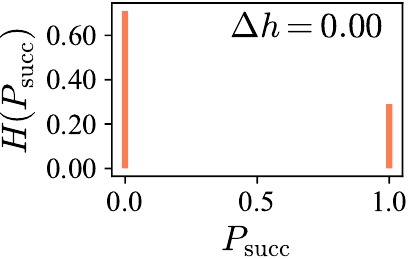}
\includegraphics[width=4.4cm]{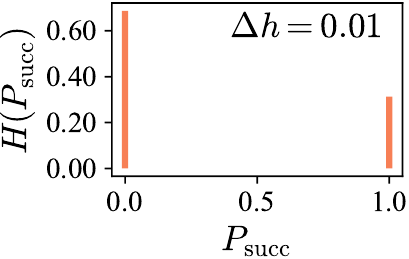}
\includegraphics[width=4.4cm]{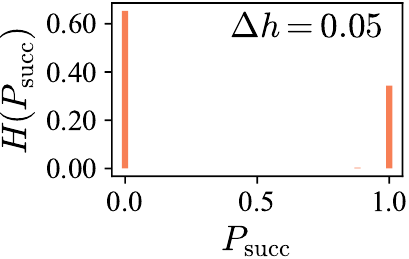}
\includegraphics[width=4.4cm]{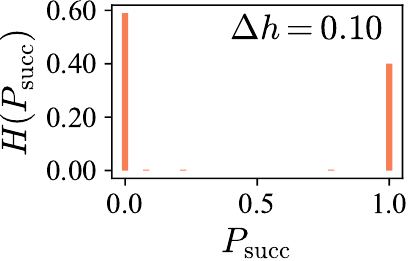}
\includegraphics[width=4.4cm]{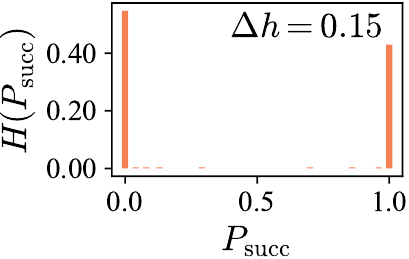}
\includegraphics[width=4.4cm]{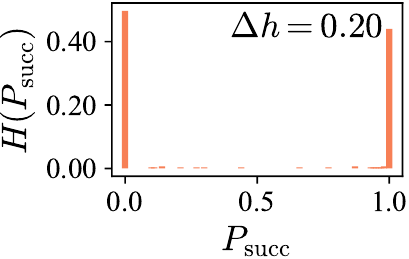}
\includegraphics[width=4.4cm]{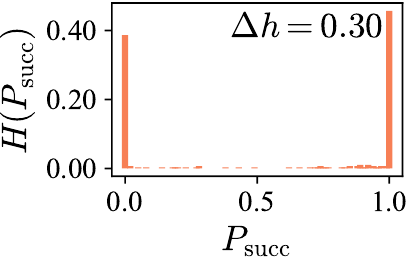}
\includegraphics[width=4.4cm]{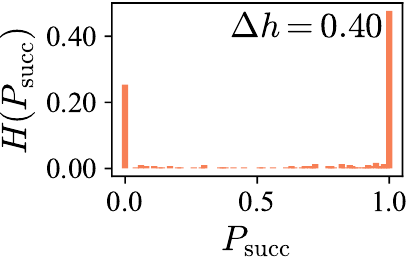}
\includegraphics[width=4.4cm]{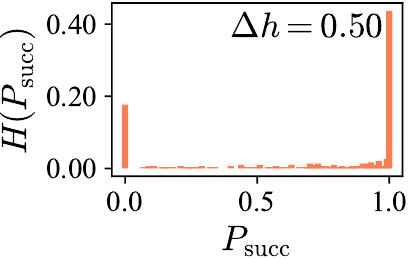}
\includegraphics[width=4.4cm]{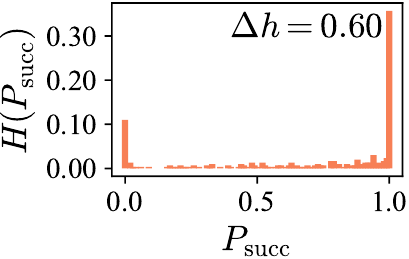}
\includegraphics[width=4.4cm]{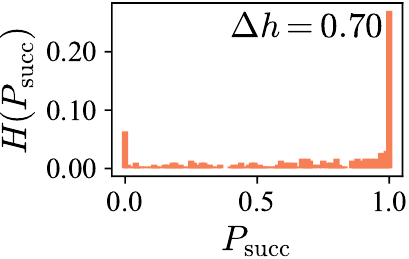}
\includegraphics[width=4.4cm]{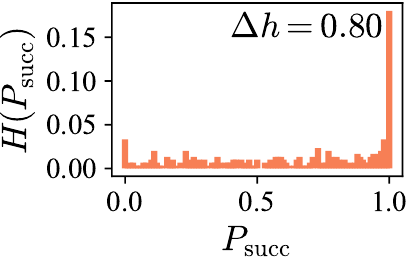}
\includegraphics[width=4.4cm]{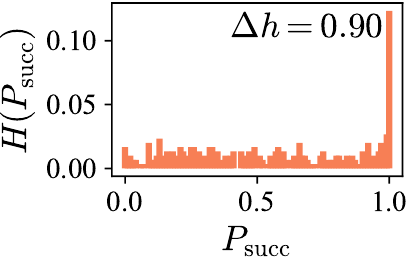}
\includegraphics[width=4.4cm]{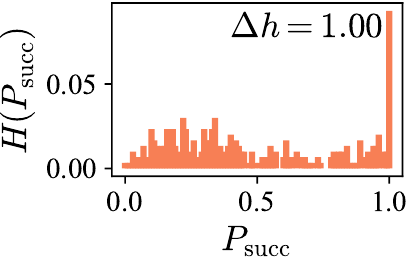}
\includegraphics[width=4.4cm]{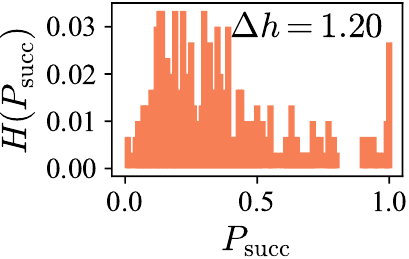}
\includegraphics[width=4.4cm]{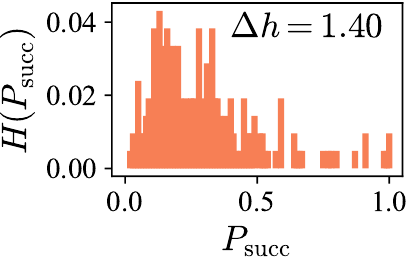}
\includegraphics[width=4.4cm]{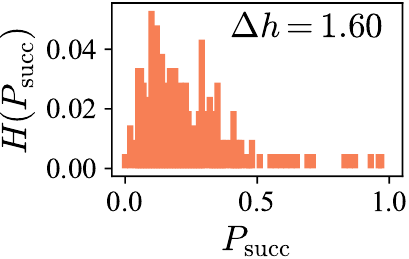}
\includegraphics[width=4.4cm]{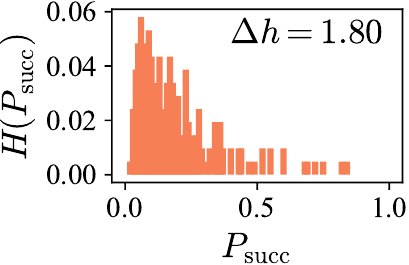}
\includegraphics[width=4.4cm]{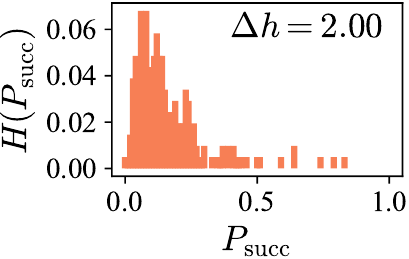}
\caption{Distributions $H(P_{\rm succ})$ for $\alpha=0.07$ and different values of $\Delta h$ as in the legends.}
\label{fig:successprobability3}
\end{figure}

The distributions $H(P_{\rm succ})$ for $\alpha=0.02,0.05,0.07$ are shown for different values of $\Delta h$ in Figs.~\ref{fig:successprobability1}, \ref{fig:successprobability2}, and~\ref{fig:successprobability3}. From these data, the average success probability is computed as $\overline P_{\rm succ}=\sum_{P_{\rm succ}}H(P_{\rm succ})P_{\rm succ}$. As evident from the figures, the distributions $H(P_{\rm succ})$ evolve as a function of $\Delta h$ in a qualitatively similar way for all the three values of $\alpha$: The distribution is bimodal at threshold, i.e., $H(P_{\rm succ})\neq0$ only for $P_{\rm succ}=0$ or $P_{\rm succ}=1$, and becomes peaked around smaller and smaller values of $P_{\rm succ}$ as $\Delta h$ increases.

\end{document}